\documentclass[pra,twocolumn,showpacs]{revtex4}

\def\>{\rangle}
\def\<{\langle}

\def\Tr{\hbox{Tr}}
\usepackage{graphicx}
\begin{document}

\title{Information-disturbance tradeoff for spin coherent state estimation}
\author{Massimiliano F. Sacchi} 
\affiliation{QUIT Quantum Information Theory Group,  
%\affiliation{
CNISM and CNR - Istituto Nazionale per la Fisica della Materia,
  Dipartimento di Fisica ``A. Volta'',  via A. Bassi 6, I-27100 Pavia, Italy} 
\homepage{http://www.qubit.it} 
\date{\today}

\begin{abstract}
  We show how to quantify the optimal tradeoff between 
  the amount of information retrieved by a quantum
  measurement in estimating an unknown spin coherent state and the 
  disturbance on the state itself, and how to derive  the 
corresponding minimum-disturbing measurement.
\end{abstract}

\pacs{03.67.-a} 

\maketitle 

\section{Introduction} The tradeoff between information retrieved from
a quantum measurement and the disturbance caused on the state of a
quantum system is a fundamental concept of quantum mechanics and has
received a lot of attention in the literature 
\cite{WH,wod,sc,sten,eng,fuchs96.pra,banaszek01.prl,fuchs01.pra,banaszek01.pra,
barnum02.xxx,gmd,ozw,mista05.pra,dema,macca,mf,cv,max,paris,buscmax}.  
Such an issue is
studied for both foundations and its enormous relevance in practice,
in the realm of quantum key distribution and quantum cryptography
\cite{key,key2}.

Quantitative derivations of such a tradeoff have been obtained in the
scenario of quantum state estimation \cite{hol,qse}. The optimal
tradeoff has been derived in the following cases: in estimating a
single copy of an unknown pure state \cite{banaszek01.prl}, many
copies of identically prepared pure qubits \cite{banaszek01.pra} and
qudits \cite{mf}, a single copy of a pure state generated by
independent phase-shifts \cite{mista05.pra}, an unknown maximally
entangled state \cite{max}, an unknown coherent state \cite{cv} and
Gaussian state \cite{paris}. Experimental realization of
minimal-disturbing measurements has been also reported \cite{dema,cv}. 
Recently, the optimal tradeoff has been also derived for quantum 
state discrimination \cite{buscmax}. 

The problem is typically the following. One performs a measurement on
a quantum state picked (randomly, or according to an assigned a priori
distribution) from a known set, and evaluates the retrieved
information along with the disturbance caused on the state. To
quantify the tradeoff between information and disturbance, one can
adopt two mean fidelities \cite{banaszek01.prl}: the estimation
fidelity $G$, which evaluates on average the best guess we can do of
the original state on the basis of the measurement outcome, and the
operation fidelity $F$, which measures the average resemblance of the
state of the system after the measurement to the original one.

In this paper, we study the optimal tradeoff between estimation and
operation fidelities when the state is a completely unknown
spin coherent state.

Our results will be obtained by exploiting the group simmetry of the
problem, which allows us to restrict our analysis on {\em covariant
  measurement instruments}. In fact, the property of covariance
  generally leads to a striking simplification of problems that may
  look intractable, and has been thoroughly used in the context of
  state and parameter estimation \cite{hol,qse}. 

The derivation of the optimal tradeoff for spin coherent state estimation 
might find application in the problem of how to 
achieve a secure distribution of a private shared
directional reference frame \cite{brs,chiri}. This task
can be achieved by setting up a secure classical key using regular
quantum key distribution, and then converting this into a private
shared reference frame using the technique of Ref. \cite{chiri}. 
However, it is conceivable that there may be some benefit to using
a different, more direct protocol wherein one sends the systems by 
encoding the directional information over the public channel and
monitors for eavesdropping upon them.  If the signal states were
$SU(2)$ coherent states, then the question of how much security
can be achieved depends on the nature of the information gain---disturbance 
tradeoff.

The paper is organized as follows.  In Sec. II we briefly review the
concept of spin coherent states. In Sec. III we show that the tradeoff
between estimation and disturbance can be studied without loss of
generality by considering measurement instruments with a covariant
symmetry with respect to the rotation group. In Sec. IV we show how to
quantify the optimal information-disturbance tradeoff and to obtain
the corresponding minimum-disturbing measurement. We close the paper in Sec. V
with concluding remarks.

\section{Spin coherent states}

In the infinite dimensional space of the harmonic oscillator states we
can construct the creation and annihilation operators, $a^\dagger,a$,
obeying the boson commutation relation $[a,a^\dagger]=1$. The coherent
states of such a system (harmonic oscillator coherent states) are
eigenvectors of the annihilation operator $a|\alpha\rangle
=\alpha|\alpha\rangle $ and can be obtained as displacements of the
ground state $|0\rangle $: 
\begin{equation} |\alpha\rangle
=D(\alpha)|0\rangle , \quad D(\alpha)= \exp(\alpha
a^\dagger-\alpha^*a).  
\end{equation} 
An important property of
such coherent states is that they satisfy the lower bound on product
of the dispersions of the position and momentum operators (or the
quadrature operators) required by the Heisenberg uncertainty relation. 

The concept of coherent states is not restricted to the infinite
dimensional space. In a finite dimensional space one can introduce
different kinds of coherent states \cite{perelomov}. In this paper we
shall concentrate on so called spin coherent states ($SU(2)$ coherent
states), which we define below. 

Let us consider the Hilbert space of spin states with total spin
$j$, hence with dimension $d=2j+1$. By $|m\rangle $,
$m=-j,-j+1,\dots,j-1,j$, we denote the basis consisting of the
eigenvectors of $J_z$ operator. Spin coherent states are defined as
rotations of the ``ground'' state $|-j\rangle$ by unitary operators
from the irreducible $SU(2)$ representation in the $2j+1$ dimensional
space: \begin{equation} |\theta, \phi\rangle = R_{\theta,
\phi}|-j\rangle , \quad R_{\theta, \phi}=e^{i\theta(J_x \sin\phi
-J_y\cos\phi)}.  \end{equation} The operator $R_{\theta, \phi}$
corresponds to a rotation by the angle $\theta$ around the axis
$\vec{n}=[\sin\phi,-\cos\phi,0]$. For $j=1/2$ the dimension of the
space is $d=2$ (qubit). In this case spin coherent states are actually
all the pure states in the space (every pure state can be described by
a direction on the Bloch sphere). In higher dimensions, however, spin
coherent states constitute only a subset in the set of all states of a
given Hilbert space, and, moreover, they approach harmonic oscillator
coherent states when the dimension of the space tends to infinity
\cite{spincoh}.

One can decompose a spin coherent state $|\theta,\phi\rangle $ in the
$J_z$-eigenvectors basis as follows \cite{perelomov} 
\begin{eqnarray} \label{eq:decomposition}
\langle m|\theta,\phi\rangle &= & 
\left( \begin{array}{c} 2j\\ j+m
\end{array}
\right)^{1/2} 
\\
& \times & 
\left(- \sin \frac \theta 2 \right )^{j+m}
\left(\cos \frac \theta 2 \right )^{j-m}\,e^{-i(j+m)\phi}\;, \nonumber 
\end{eqnarray}

A spin coherent state in a Hilbert space ${\cal H}$, with
$\hbox{dim}({\cal H})=2j+1$ will be written as $|\psi _g \rangle =
U_g |-j \rangle $, where $U_g $ is a unitary irreducible representation of the 
$SU(2)$ group in dimension $d $ \cite{nota1}. When performing averages
on group parameters, for convenience we will take the normalized
invariant Haar measure $dg$ over the group, i.e. $\int_{SU(2)} dg=1$,
and we will also omit $SU(2)$ from the symbol of integral.

\section{Covariant instruments for the rotation group}
A measurement process on a quantum state $\rho $ with outcomes $\{r
\}$ is described by an {\em instrument} \cite{instr}, namely a set of
trace-decreasing completely positive (CP) maps $\{{\cal E}_r \}$. Each
map is a superoperator that 
%can then be written in the Kraus form \cite{kraus}
%\begin{eqnarray}
%{\cal E}_r( \rho )= \sum _\mu  A_{r \mu} \rho A_{r \mu} ^\dag \;,
%\label{uno}
%\end{eqnarray}
%and 
provides the state after the measurement 
\begin{eqnarray}
\rho _r =\frac{{\cal E}_r
[\rho ]}{\Tr \{{\cal E}_r [\rho ]\} }
\;,
\end{eqnarray}
along with the probability of
outcome 
\begin{eqnarray}
p_r = \Tr \{{\cal E}_r [\rho ]\} 
%= \Tr\left [\sum _\mu A^\dag _{r
%  \mu}A_{r\mu } \rho\right ]
\;.
\end{eqnarray}
The set of positive operators $\{ \Pi _r = {\cal E}_r ^\vee [I] \}$, 
%\sum _\muA^\dag _{r \mu}A_{r\mu }\}$ 
where $\vee $ denotes the dual map satisfying $\Tr \{{\cal E}^\vee [A] \,B\}
=\Tr \{A \,{\cal E}[B]\}$ for all $A$ and $B$, 
is known as positive operator-valued
measure (POVM), and normalization requires the completeness relation
$\sum _r {\cal E}^\vee _r = I$. This is equivalent to require that the map $\sum
_r {\cal E}_r $ is trace-preserving. In the following we will denote a 
$SU(2)$ unitary map as ${\cal U}_g$, namely ${\cal U}_g [\rho ]=
U_g \rho U^\dag _g$. Notice that ${\cal U}^\vee _g= {\cal U}_{g^{-1}}$. 

\par The operation fidelity $F$ evaluates on average how much the
state after the measurement resembles the original one, in terms of
the squared modulus of the scalar product.  Hence, for a measurement
of an unknown spin coherent state, one has
\begin{eqnarray}
F = \int dg \sum _{r }  
\langle -j | {\cal U}_g^\vee \circ {\cal E}_r \circ {\cal U}_g 
[|-j\rangle \langle -j |] |-j \rangle \;,\label{fm}
\end{eqnarray}
By adopting a guess function $f$, 
for each measurement outcome $r$ one guesses
a spin coherent states $|\psi _ {f(r)} \rangle $, and the
corresponding average estimation fidelity is given by 
\begin{eqnarray}
G
&=&
\int  dg \sum _{r}  \Tr\{ {\cal E}_r \circ {\cal U}_g [|-j \rangle\langle -j|]
\}  
\nonumber \\& \times & 
\langle \psi_{f(r)}|{\cal U}_g [|-j \rangle\langle -j|]|\psi _{f(r)}\rangle 
\;.\label{gm}
\end{eqnarray}
We are interested in the optimal tradeoff between $F$ and $G$, and 
without loss of generality we can restrict our attention to {\em
  covariant} instruments, that satisfy 
\begin{eqnarray}
{\cal U}^\vee _g \circ {\cal E} _h \circ {\cal U}_g ={\cal E}_{g^{-1}h}
\;.\label{cvv}
\end{eqnarray}
In fact, for any instrument $\{ {\cal E}_r \}$ and guess function $f$, 
it turns out that  
the covariant instrument
\begin{eqnarray}
\widetilde {\cal E}_h =
\sum _{r } 
{\cal U}_h \circ {\cal U}^\vee  _{f(r)} 
\circ {\cal E}_r \circ {\cal U} _{f(r)} \circ {\cal U}^\vee  _h 
\;\label{cove} 
\end{eqnarray} 
with continuous outcome $h\in SU(d)$, 
along with the guess $|\psi _h \rangle $, provides the
same values of $F$ and $G$ as the original instrument $\{{\cal E}_r\}$. Moreover, 
for covariant instruments the optimal guess function automatically 
turns out to be the identity function. 

It is useful now to consider the Jamio\l kowski representation \cite{CJ,max2}, 
that gives a one-to-one correspondence between a CP map ${\cal E}$
from ${\cal H}_{in}$ to ${\cal H}_{out}$ and a
positive operator $R$ on ${\cal H}_{out}\otimes {\cal H}_{in}$ 
through the equations 
\begin{eqnarray}
&&{\cal E}(\rho )=\Tr _{in}[(I_{out} \otimes \rho ^\tau ) R ]\;,
\nonumber \\& & 
R=({\cal E }\otimes I_{in}) |\Phi \rangle \langle \Phi | \;,\label{jam}
\end{eqnarray}
where $|\Phi \rangle $ represents the (unnormalized) maximally entangled state of 
${\cal H}_{in}\otimes {\cal H}_{in}$, and $\tau $ denotes the transposition on the fixed basis.  
When ${\cal E}$ is trace preserving, correspondingly one has $\Tr
_{out}[R] = I_{in}$. 
For covariant instruments ${\cal E}_g$ as in Eq. (\ref{cvv}), the
operator $R_g$ acts on ${\cal H}^{\otimes 2}$, and has the form
\begin{eqnarray} 
R_g = U_g \otimes U_g ^{*} R_0 U_g ^{\dag }\otimes
U_g ^{\tau } \;,\label{rggg} 
\end{eqnarray} 
(where $*$ denotes complex conjugation) with $R_0 \geq 0$, and the
trace-preserving condition 
\begin{eqnarray} \int dg \, \Tr _{2}[R_g] =
I \;.\label{34} 
\end{eqnarray} 
From Eq. (\ref{rggg}) and the identity
(Schur's lemma for irreducible group representations \cite{zelo})
\begin{eqnarray} \int dg \,U_g X U_g^\dag= \frac {\Tr[X]}{\Tr[ I]} I
\label{eq:gravtrc}\;, \end{eqnarray} 
it follows that condition
(\ref{34}) is equivalent to 
\begin{eqnarray} \Tr [R_0]=2j+1
\;. \label{tr134} \end{eqnarray} 

\section{Optimal information-disturbance tradeoff}
For covariant instruments, the expressions of the fidelities 
$F$ and $G$ in Eqs. (\ref{fm}) and (\ref{gm}) can be rewritten as follows
\begin{eqnarray}
F&=& \int dg \int dh \, 
\langle -j | {\cal U}_g^\vee \circ {\cal E}_h \circ {\cal U}_g 
[|-j\rangle \langle -j |] |-j \rangle 
\nonumber \\
&=& \int dg \, 
\langle -j |{\cal E}_g [|-j \rangle \langle -j |]|-j \rangle\;,\\
G&=& \int dg \int dh \, 
\Tr\{ {\cal E}_h \circ {\cal U}_g [|-j \rangle\langle -j|]
\}  
\nonumber \\& \times & 
\langle -j| {\cal U}^\vee _h \circ {\cal U}_g 
[|-j \rangle\langle -j|]|-j\rangle 
\nonumber \\
&=& \int dg \,
\Tr\{ {\cal E}_g [|-j \rangle\langle -j|]\}
\nonumber \\& \times & 
\langle -j| {\cal U}_g 
[|-j \rangle\langle -j|]|-j\rangle   
\;,
\end{eqnarray}
where the covariance property (\ref{cvv}) and the invariance of the 
Haar measure have been used. 
Moreover, using the isomorphism (\ref{jam}), we can write $F$ and $G$ as 
$F=\Tr [R_F R_0]$ and $G=\Tr [R_G R_0]$, where $R_F$ and
$R_G$ are the following positive operators 
\begin{eqnarray} R_F &=&
\int dg \,U_g \otimes U_g ^{*} \,|-j \rangle \langle -j | ^{\otimes
2}\, U_g ^{\dag }\otimes U_g ^{\tau } \;,\label{rff} \\ 
%\end{eqnarray}
%and %\begin{eqnarray} 
R_G&=& \int dg \, |\langle -j | U_g |-j \rangle
|^2\, 
%\times \nonumber \\& & 
I \otimes (U_g ^{*} \,|-j \rangle \langle
-j | \, U_g ^{\tau } )\nonumber \\& =& I \otimes \Tr _{1} [(|-j 
\rangle \langle -j  |\otimes I )  R_F] 
\;.\end{eqnarray}
Using Schur's lemma for reducible group representations \cite{zelo},
one can evaluate the group integral in Eq. (\ref{rff}) from the identity  
\begin{eqnarray} 
&&\int dg \,U_g \otimes U_g ^* \,Y \, U_g^{\dag }\otimes U_g ^\tau 
\nonumber \\ 
&& = \left (\int dg \,U_g \otimes U_g \, Y ^\theta \, U_g^{\dag
}\otimes U_g ^\dag \right )^{\theta }
\nonumber \\ 
&&= \sum_{l=0}^{2j} \Tr[Y^\theta P_{l}]
\frac{P_{l}^\theta }{\Tr{P_{l}}}\;,
\end{eqnarray}
where $\theta $
denotes the partial transpose on the second Hilbert space, and $P_{l}$
represents the projector on the subspace of ${\cal H}\otimes {\cal H}$
with total spin $l$.
Then, one has 
\begin{eqnarray}
R_F 
&= & 
\frac{1}{4j+1} P_{2j}^\theta\;,
\\
R_G &= &  
\frac{1}{4j+1} I \otimes \Tr _1 [(|-j \rangle \langle -j | \otimes I ) P_{2j}]
\;.
\label{rgg}
\end{eqnarray}
The optimal tradeoff between $F$ and $G$ can be found by looking for a
positive operator $R_0$ that satisfies the trace-preserving condition 
(\ref{tr134}) and 
maximizes a convex combination 
\begin{eqnarray}
p G +(1-p) F = 
 \Tr \{[pR_G +(1-p)R_F ]R_0\} \;, \label{cx}
\end{eqnarray}
where $p \in [0,1]$ controls the tradeoff between the quality of the
state estimation and the quality of the output replica of the
state. Then, $R_0$ will provide a covariant instrument that
achieves the optimal tradeoff.  
In fact, we are interested in maximizing the operation fidelity
$F=\Tr[R_F R_0]$, for a fixed value of the estimation fidelity
$G=\Tr[R_G R_0]$. This is equivalent to maximizing the convex
combination (\ref{cx}). Indeed, 
suppose that for a given value of $p$, we find $R_0$ that
maximizes (\ref{cx}).  It is clear that for $G=\Tr[R_G R_0]$ this map
yields maximum possible $F$, because any higher $F$ would increase (\ref{cx}). 

It turns out that for any $p$ the
eigenvector corresponding to the maximum eigenvalue of $C(p)\equiv p
R_G +(1-p)R_F$ is non degenerate and of
the form \cite{nota2}
\begin{eqnarray}
|\chi \rangle = \sum _{n=-j}^j a_n  |n \rangle |n\rangle 
\;,\label{frm}
\end{eqnarray}
with suitable positive $\{a_n \}$. Upon taking $R_0 $ proportional to
$|\chi \rangle \langle \chi |$ and satisfying (\ref{tr134}),  
the corresponding covariant instrument will then be
optimal. 

Notice that one has 
\begin{eqnarray}
&& F_{min}\equiv \frac{2j+1}{4j+1}\leq F \leq 1 \;,\\
&& G_{min}\equiv \frac{1}{2j+1}
\leq G \leq \frac{2j+1}{4j+1}\equiv
G_{max} \;,
\end{eqnarray}
where $G_{max }$ is the optimal estimation fidelity with
corresponding operation fidelity $F_{min}$, and $G_{min}$ corresponds
to the value of $G$ for a random guess of the unknown state. The values 
$G_{max}$, $F_{min}$ for the optimal estimation are obtained for 
$R_0 = (2j+1) |-j \rangle \langle -j| ^{\otimes 2}$, corresponding to a 
quantum measurement described by spin coherent state POVM, i.e.
\begin{equation}
{\cal E}_g (\rho) = (2j+1) |\psi _g \rangle \langle \psi _g | \rho  
|\psi _g \rangle \langle \psi _g |\;.\label{opmap}
\end{equation}
On the other hand, the values $F=1$ and $G=G_{min}$ are obtained for  
$R_0 = (2j+1) \sum _{n,m=-j}^j |n,n \rangle \langle m, m|$, 
which corresponds to the 
identity operation. 

Once one recognizes that the eigenvector $|\chi \rangle $ is of 
the form as in Eq. 
(\ref{frm}), the optimization problem can be rewritten as
\begin{eqnarray}
F= \frac{1}{4j+1} \,\max _{\{a_n \}} 
\sum _{n,m=-j}^j a_n a_m c^2_{n,m}
\end{eqnarray}
with the constraints
\begin{eqnarray}
\frac{1}{4j+1} 
\sum _{n=-j}^j a_n^2  c_{n,-j}^2 =G \;, \qquad \sum _{n=-j}^j a_n ^2 =2j+1 \;, 
\end{eqnarray}
where $c_{n,m}$ denotes Clebsch-Gordan coefficients $c_{n,m}=
\langle j,n;j,m |(2j),n+m \rangle$.  
Notice that from the property 
\begin{eqnarray}
&&\langle j_1,n;j_2,m |j,n+m \rangle = \nonumber \\
&&
(-1)^{j_1+j_2-j}
\langle j_2,m;j_1,n |j,n+m \rangle 
\end{eqnarray}
it follows that the matrix $(c_{n,m})$ is symmetric. One can solved numerically such a 
constrained maximization, thus obtaining the optimal tradeoff between the 
operation and the estimation fidelities, along with the corresponding optimal 
measurement instrument
\begin{equation}
{\cal E}_g (\rho )=U_g \xi U_g^\dag \,\rho U_g \xi U_g ^\dag \;,\label{sol}
\end{equation} 
with $\xi =\sum _{n=-j}^j a_n  |n \rangle \langle n |$. 

For example, in the histogram of Fig. 1, we plot the optimal
coefficients $\{a_n \}$ for the minimum-disturbing measurement of a
spin coherent state with $j=2$ and fixed estimation fidelity $G=1/2$,
for which the maximum value of the operation fidelity $F \simeq 0.795$
is achieved.  

\begin{figure}[htb]
\begin{center}
\includegraphics[scale=.6]{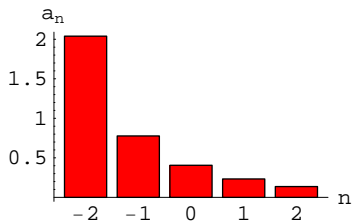}
\caption{Optimal coefficients $\{a_n \}$ that provide through Eq. (\ref{sol}) 
the minimum-disturbing measurement for the estimation 
of a spin coherent state with $j=2$ and fixed estimation fidelity $G=1/2$, 
for which the maximum value of the operation fidelity  
$F \simeq 0.795$ is achieved.}
\label{f:fig1}
\end{center}
\end{figure}

\par We can introduce two normalized quantities that can be
interpreted as the average information $I$ retrieved from the quantum
measurement and the average disturbance $D$ affecting the original
quantum state as follows: 
\begin{eqnarray} I=\frac{G -
G_{min}}{G_{max}-G_{min}} \;,\label{igg} \end{eqnarray} 
and
\begin{eqnarray} D= \frac{1-F}{1-F_{min}} \;.\label{dgg}
\end{eqnarray} 
Clearly, one has $0\leq I \leq 1$, and $0\leq D \leq
1$.

\begin{figure}[hbt]
\begin{center}
\includegraphics[scale=.7]{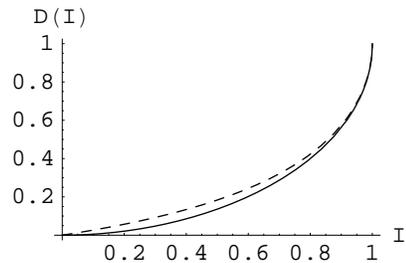}
\caption{Optimal information-disturbance tradeoff in estimating an
  unknown spin coherent state for $j=1$ (solid line), and a harmonic oscillator 
coherent state (dotted line), where $I$ and $D$ are
  defined through Eqs. (\ref{igg}) and (\ref{dgg}) in terms of the
  estimation and operation fidelities $G$ and $F$, respectively. For
  given value of the retrieved information $I$, the curve $D(I)$ are
  a lower bound for the disturbance of any measurement instrument.}
\label{f:fig2}
\end{center}
\end{figure}

In Fig. 2 (solid line) we plot the  optimal information-disturbance tradeoff, for $j=1$.
The curve $D(I)$ represents a lower bound for the disturbance of any measurement 
instrument that gathers information I. The bound is achieved by a covariant 
instrument as in Eq. (\ref{sol}). 
The optimal tradeoff $D(I)$ depends very slighly  on the value 
of the spin $j$. In fact, it is known that for $j \to \infty$, spin coherent states 
approach the standard coherent states of harmonic oscillator \cite{spincoh}.  
From Eq. (5) of Ref. \cite{cv}, for harmonic oscillator coherent states 
one can obtain the following expression for the optimal information-disturbance tradeoff 
\begin{equation}
D(I)=1- \sqrt{\frac{2(1-I)}{2-I}}\,,
\end{equation}
which has been plotted in dotted line in Fig. 2.

We can consider the quantity $\Tr [\xi ] =\sum _{n=-j}^j a_n$ as  a 
global quantity that characterizes the measurement instrument of Eq. (\ref{sol}) 
achieving the optimal tradeoff. 
In fact, using  Jensen's inequality,  
one has 
\begin{equation}
\sqrt{2j+1}\leq \Tr [\xi ] \leq 2j +1 \;, 
\end{equation}
where the lower and upper bound correspond to the optimal estimation map 
(\ref{opmap}) and the identity map (with no information neither disturbance), 
respectively. Notice that $\Tr [\xi ]$ is related to the projection of 
the optimal bipartite vector $R_0$ on the maximally entangled vector 
$|\Phi \rangle =\sum_{n=-j}^j |n \rangle |n \rangle$ 
by the relation $\Tr [\xi ]= 
\sqrt {\langle \Phi |R_0|\Phi \rangle}$.   
In Fig. 3, we plot the value of $\Tr [\xi ]$ for the 
minimum-disturbing measurement versus the information, for spin $j=1$.

\begin{figure}[hbt]
\begin{center}
\includegraphics[scale=.7]{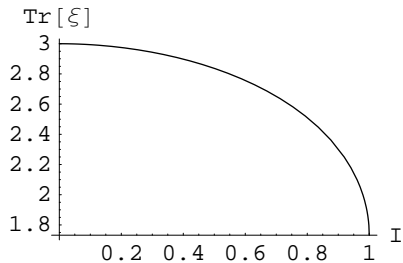}
\caption{The quantity $\Tr[\xi]$ for the measurement 
instrument of Eq. (\ref{sol})  
that achieves the optimal information-disturbance tradeoff versus 
the information $I$, 
for spin $j=1$.} 
\label{f:fig3}
\end{center}
\end{figure}

\section{Conclusions} In conclusion, we have shown how to derive the
optimal tradeoff between the quality of estimation of an unknown spin
coherent state and the degree the initial state has to be changed by
this operation. The optimal tradeoff can be achieved by a covariant
measurement instrument as in Eq. (\ref{sol}). By suitable
normalization of the estimation and operation fidelities, the optimal
tradeoff is shown to be almost independent of the value of the spin
$j$. 
\par In the case of estimation of an unknown
pure state \cite{banaszek01.prl} or maximally entangled state
\cite{max}, the minimum-disturbing
measurement is simply the ``coherent superposition'' of the
measurement instrument for optimal estimation and the identity map, 
namely the Kraus operators achieving the optimal information-disturbance 
tradeoff are just the sum of those corresponding to maximum 
information extraction and minimum disturbance. For spin coherent state, 
the solution is more complex.  
This is due to the fact that the derivation of the tradeoff in a 
covariant estimation problem involves the 
evaluation of group integrals as in $R_F$ and $R_G$.   
Generally, such integrals give a sum 
of $N$ operators, where $N$ is the number of invariant subspaces 
of the representation $U_g \otimes U_g^*$, 
and the optimal Kraus operators are the sum of $N$ corresponding terms. 
For pure states or maximally entangled states,  
$U_g\in SU(d)$ in dimension $d$, and $N=2$ 
(the symmetric and antisymmetric subspaces). For spin coherent states,  
$U_g$ is a unitary irreducible representation of $SU(2)$ in dimension
$2j+1$, and the invariant subspaces of $U_g \otimes U_g^*$ are $N=2j+1$. 
Correspondingly, the Kraus operators of the 
minimum-disturbing measurement are given by a sum of $2j+1$ operators.  

\section*{Acknowledgments} 
Stimulating discussions with G. Chiribella and P. Perinotti are acknowledged. 
This work has been sponsored by Ministero 
Italiano dell'Universit\`a e della Ricerca (MIUR) through FIRB (2001)
and PRIN 2005.


\begin{thebibliography}{99}
\bibitem{nota1} More precisely, a spin coherent state corresponds to a point of the 
two-dimensional sphere $S^2=SU(2)/U(1)$, where the quotient group 
$SU(2)/U(1)$ takes into account the stability group $U(1)$ corresponding to the rotations 
around the $J_z$-axis, which leave the state $|-j\rangle $ invariant, a part from irrelevant phase factor.
\bibitem{nota2}Such a result can be checked, for example, 
by means of the power method \cite{pm}.
\end{thebibliography}
\end{document}